\documentclass[3p,times]{elsarticle}

\usepackage{ecrc}

\volume{00}

\firstpage{1}

\journalname{Nuclear Physics A}

\runauth{}

\jid{npa}

\jnltitlelogo{Nuclear Physics A}

\usepackage{amssymb}

\biboptions{square,comma,numbers,sort&compress}

\usepackage[figuresright]{rotating}

\newcommand{\pp}{\rm pp}
\newcommand{\sqrts}{\sqrt{s}}
\newcommand{\sqrtsNN}{\sqrt{s_{\rm NN}}}

\newcommand{\gev}{\mathrm{GeV}}
\newcommand{\tev}{\mathrm{TeV}}

\newcommand{\mum}{\mathrm{\mu m}}

\newcommand{\ptrans}{p_{\rm T}}

\newcommand{\DtoKpi}{{\rm D^0\to K^-\pi^+}}
\newcommand{\DtoKpipi}{{\rm D^+\to K^-\pi^+\pi^+}}
\newcommand{\DstartoDpi}{{\rm D^{*+}\to D^0\pi^+}}
\newcommand{\DsToPhipi}{{\rm D_{\rm s}^{+}\to \phi \pi^+}}
\newcommand{\Dzero}{{\rm D^0}}
\newcommand{\Dstar}{{\rm D^{*+}}}
\newcommand{\Dplus}{{\rm D^+}}
\newcommand{\Ds}{{\rm D_{\rm s}^{+}}}

\newcommand{\vtwo}{v_{2}}
\newcommand{\RAA}{R_{\rm AA}}
\newcommand{\RpPb}{R_{\rm pPb}}
\newcommand{\RAAD}{R_{\rm AA}^{\rm D}}
\newcommand{\RAAfd}{R_{\rm AA}^{\rm feed-down}}
\newcommand{\RpPbfd}{R_{\rm pPb}^{\rm feed-down}}
\newcommand{\RpPbD}{R_{\rm pPb}^{\rm D}}
\newcommand{\RAAB}{R_{\rm AA}^{\rm B}}

\newcommand{\jpsi}{{\rm J/}\psi}

\newcommand{\avTAA}{\langle T_{\rm AA} \rangle}

\begin{document}

\begin{frontmatter}



\dochead{}

\title{Measurement of D-meson production in pp, p--Pb and Pb--Pb collisions with ALICE at the LHC}


\author[arossi]{A. Rossi}
\author{for the ALICE Collaboration.}

\address[arossi]{CERN}

\begin{abstract}
Heavy quarks, i.e. charm and beauty, are considered calibrated probes for the strongly interacting
deconfined medium (Quark Gluon Plasma, QGP) formed in heavy-ion collisions. Produced in hard
scattering processes in the initial stages of the collision, they interact with the medium, lose energy
and, depending on the coupling strength to the medium, take part in the collective motion of the
QCD matter. ALICE measured the production of $\Dzero$, $\Dstar$, $\Dplus$ and $\Ds$ mesons at central rapidity in
pp, p--Pb and Pb--Pb collisions at the LHC. The study of the modification of the transverse momentum
differential yields of charm particles in Pb--Pb collisions with respect to pp collisions,
quantified by the nuclear modification factor ($\RAA$), can unravel details of the energy loss
mechanism, such as its dependence on the quark mass and on the path length the parton travels
through the medium. A similar comparison between pp and p--Pb collision data ($\RpPb$) is
fundamental to disentangle effects related to the presence of the hot medium from cold nuclear
matter effects. The degree of thermalization
and coupling to the medium is investigated in semi-peripheral Pb-Pb collisions by measuring the
elliptic flow coefficient ($v_{2}$) at low $\ptrans$. At high $\ptrans$, $v_{2}$ is sensitive to the path-length
dependence of the energy loss. Results on the transverse momentum and centrality dependence of the D-meson elliptic flow and
$\RAA$ will be presented. The comparison with the $\RAA$ of non-prompt $\jpsi$ from B-meson decays measured
with CMS will be discussed. The preliminary results on D-meson $\RpPb$ and the dependence of D-meson
yields on rapidity in p--Pb collisions will be shown. As an outlook, the analysis and the
preliminary results on the azimuthal correlations of D-mesons and charged hadrons in pp collisions
will be described.
\end{abstract}


\end{frontmatter}



\section{Introduction}
\label{sec:intro}
The comparison of open heavy-flavour hadron production in proton-proton, proton-Pb and Pb--Pb collisions
at the LHC offers the opportunity to investigate the properties of the high-density
colour-deconfined state of strongly-interacting matter (Quark Gluon Plasma, QGP) that is expected to be produced
in high-energy collisions of heavy nuclei~\cite{pbmJSNature}. Due to their large mass, charm and beauty quarks are 
created at the initial stage of the collision in hard-scattering processes with high virtuality ($Q^{2}\gtrsim 4 {\rm m}_{c[b]}^{2}$)
involving partons of the incident nuclei. 
They interact with the medium and lose energy 
via both inelastic (medium-induced gluon radiation, or radiative energy loss)~\cite{gyulassy,bdmps} 
and elastic (collisional energy loss)~\cite{thoma} processes. The loss of energy, sensitive 
to the medium energy density and size, is expected to depend on the quark mass and be smaller for heavy quarks than
for light quarks and gluons for most of the mechanisms considered in theoretical models. In particular, the parton and mass dependence
of radiative energy loss derives from the smaller colour coupling factor of quarks with respect to gluons, 
and from the `dead-cone effect', which reduces small-angle gluon radiation for heavy quarks with moderate 
energy-over-mass values~\cite{deadcone}. A sensitive observable is the nuclear modification factor,
defined as $\RAA(\ptrans)=\frac{{\rm d}N_{\rm AA}/{\rm d}\ptrans}{\avTAA {\rm d}\sigma_{\pp}/{\rm d}\ptrans}$,
where $N_{\rm AA}$ is the 
yield measured in heavy-ion collisions, $\avTAA$ is the average nuclear overlap function calculated with the 
Glauber model~\cite{glauber} in the considered centrality range, and $\sigma_{\pp}$ is the production cross section 
in pp collisions. In-medium energy loss determines a suppression, $\RAA<1$, of hadrons at 
moderate-to-high transverse momentum ($\ptrans\gtrsim 2~\gev/c$). 
The dependence of the energy loss on the parton nature (quark/gluon) and mass can be investigated by comparing the nuclear modification factors 
of hadrons with charm $(\RAAD)$ and beauty $(\RAAB)$ with that of pions $(\RAA^{\pi})$, mostly originating
from gluon fragmentation at LHC energies. A mass ordering pattern $\RAA^{\pi}(\ptrans)<\RAAD(\ptrans)<\RAAB(\ptrans)$ 
has been predicted~\cite{deadcone,Armesto:2005iq}. However, it is important to note that the comparison 
of heavy-flavour hadron and pion $\RAA$ cannot be interpreted directly as a comparison of charm, beauty, and 
gluon energy losses, due to the different parton fragmentation functions and slope of the $\ptrans$-differential cross
sections (even in the absence of medium effects). Moreover, at low $\ptrans$, a
significant fraction of pions does not come from hard-scattering processes. 

A $\RAA$ value different from unity can also originate from initial and final state ``cold-nuclear matter'' effects, 
not related to the formation of a deconfined medium. At LHC energies, nuclear shadowing, which 
reduces the parton density for gluons carrying a nucleon momentum fraction $x$ below $10^{-2}$, %
is expected to be the most important for heavy-flavour production. A correct interpretation of heavy-ion results
demands for the measurement of these effects via the analysis of p--Pb data. 

In heavy-ion collisions with non-zero impact parameter the interaction region exhibits 
an azimuthal anisotropy with respect to the reaction plane ($\Psi_{\rm RP}$) defined
by the impact parameter and the beam direction. Collective effects
convert this geometrical anisotropy into an anisotropy in momentum space that
is reflected in the final state hadron azimuthal distribution~\cite{vtwoOllitrault}. The effect, sensitive to the degree of thermalization of the system,
can be evaluated by measuring the $2^{\rm nd}$ coefficient 
of the Fourier expansion of the particle azimuthal distribution, called elliptic
flow ($\vtwo$). The measurement of D-meson $\vtwo$
can provide, at low $\ptrans$, fundamental information on the degree of thermalization 
of charm quarks in the medium. 
At high $\ptrans$, a non-zero $\vtwo$ can originate from the path-length dependence
of energy loss~\cite{bamps,HFurQMD,CollLPMrad,powlang,whdg,TamuRappEtAl}. %

The measurement of both "hot" and "cold" nuclear effects requires the understanding of the production cross-sections of open heavy-flavour in pp collisions,
used as a reference. The $\ptrans$-differential production cross sections of D mesons is well described by fix-order 
pQCD calculations relying on the collinear factorization approach, like FONLL~\cite{fonll}, GM-VFNS~\cite{gmvfns}, or the $k_{\rm T}$-factorization 
approach~\cite{ktfact}. The study of the azimuthal 
angular correlations of D-meson and charged hadrons produced in the collisions offers the opportunity
of investigating charm production in a more differential way and can serve as a test for Monte Carlo generators simulating 
the full kinematic of charm production processes and parton shower. 

%
%
%
In these proceedings, the measurements of D-meson production in pp, p--Pb and Pb--Pb collisions performed by the ALICE Collaboration are presented.
%
%
\begin{figure*}
  \centering
  \includegraphics[height=0.25\textheight]{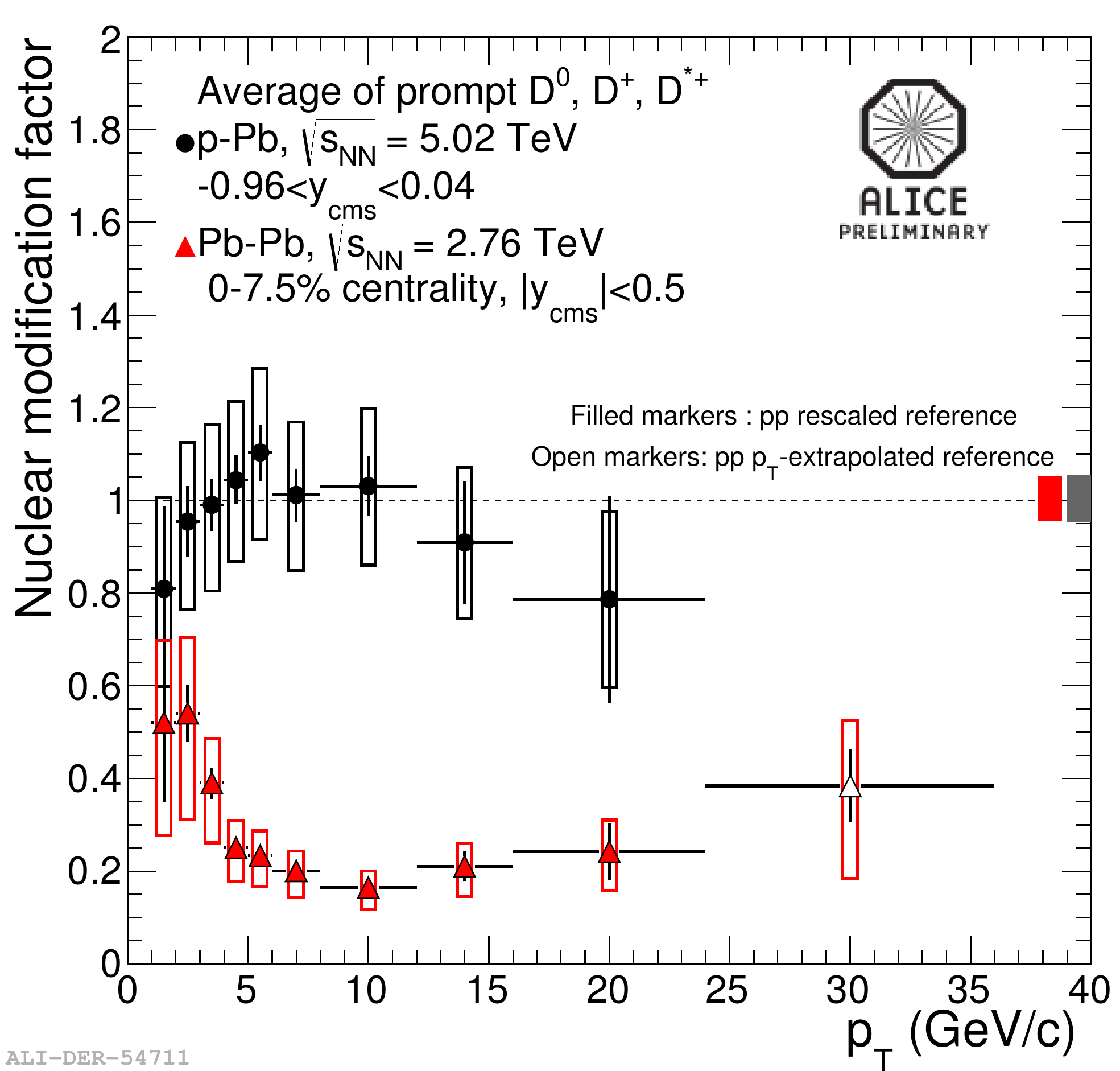}
  \includegraphics[height=0.25\textheight]{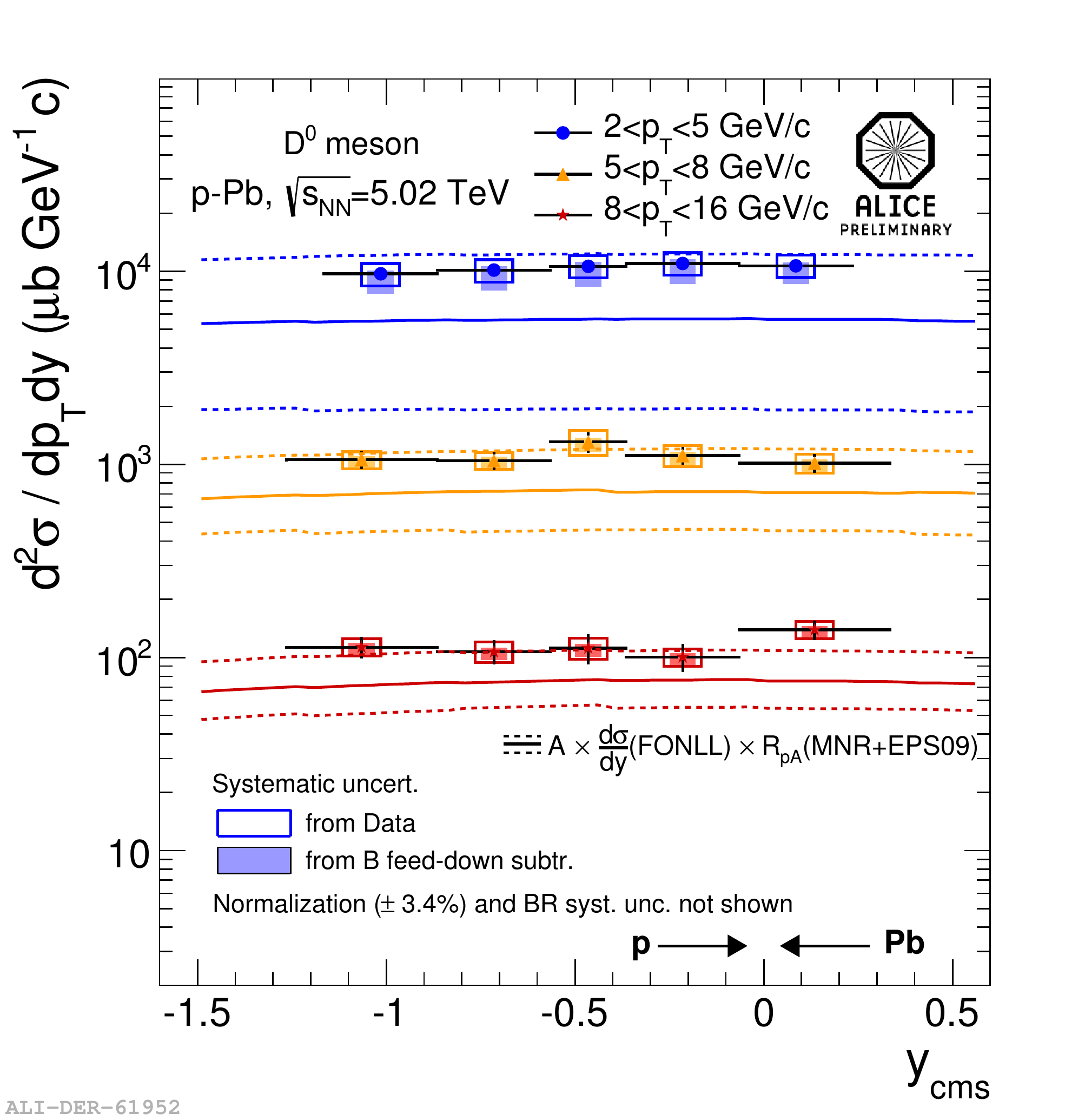}
  \caption{Left panel: comparison of average $\Dzero$, $\Dstar$ and $\Dplus$ nuclear modification factors measured in p--Pb 
    collisions and in the 0-7.5\% most central Pb--Pb collisions. Right panel: $\ptrans,y$-differential cross-section
  for $\Dzero$ production in p--Pb collisions as a function of the rapidity in the centre of mass system
  for three different $\ptrans$ ranges. The continuous and dashed lines represent expectations based on pQCD calculations
  including EPS09 parametrization of nuclear PDF~\cite{fonll,mnr,epsZeroNine} (see text for more details).} 
  \label{fig:RpPb}       
\end{figure*}
\begin{figure*}
  \centering
  \includegraphics[height=0.25\textheight]{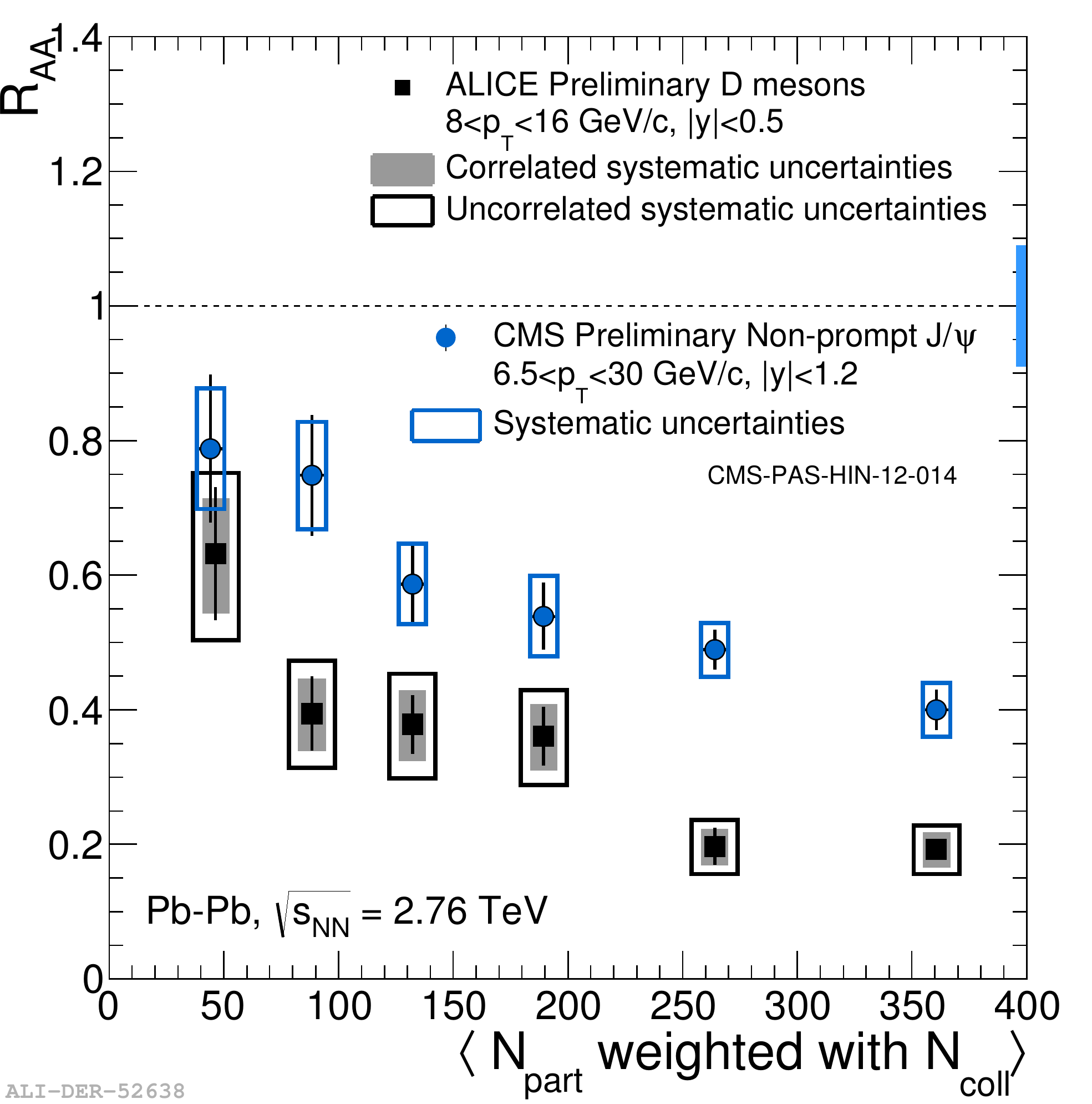}
  \includegraphics[height=0.25\textheight]{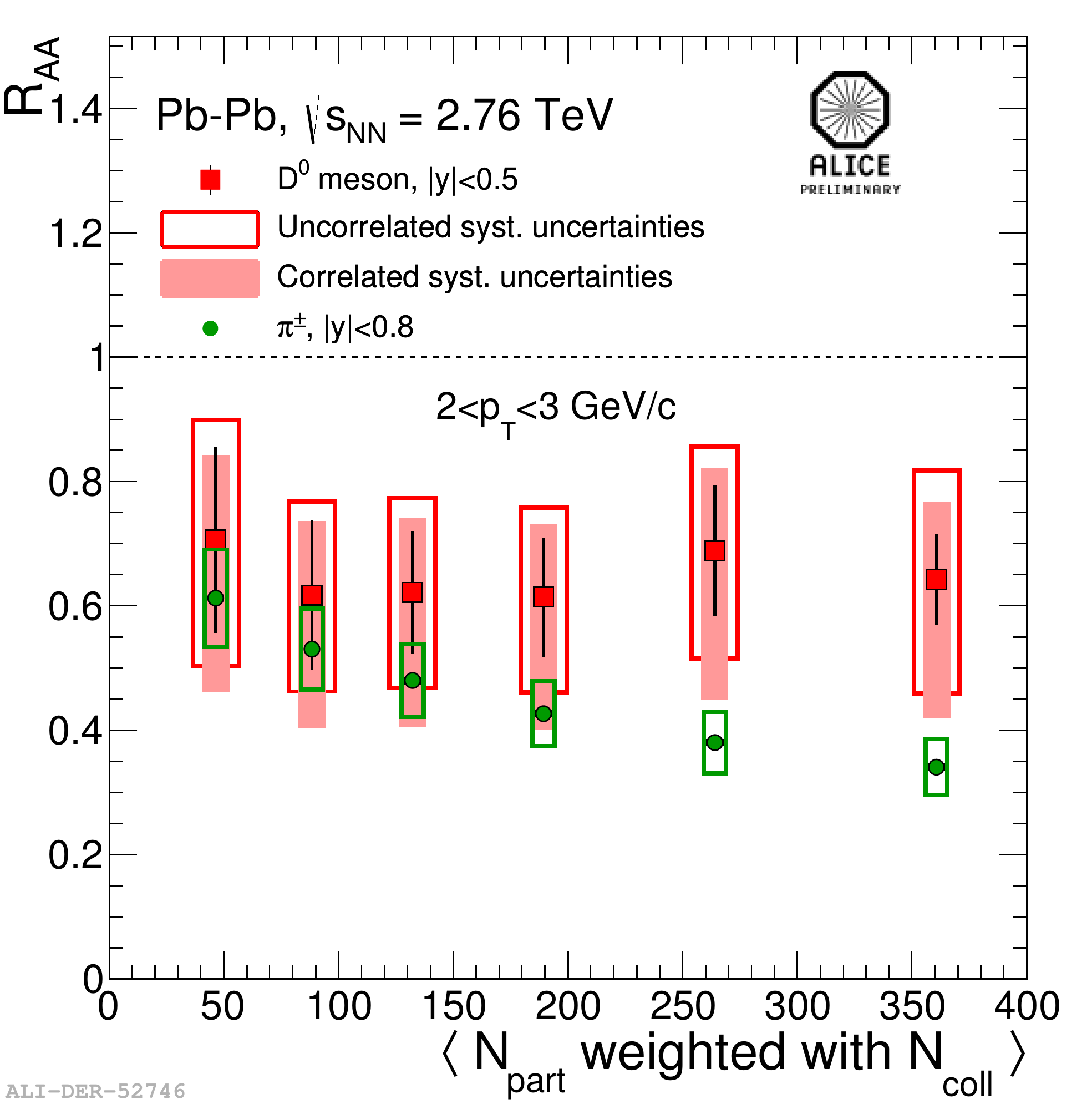}
  \caption{Comparison of the centrality dependence of the nuclear modification factors of 
    non-strange D mesons in $8<\ptrans<16~\gev/c$ and $\jpsi$ from B-meson decays measured with
    CMS~\cite{CMSjpsiPAS} in $6<\ptrans<30~\gev/c$ (left panel) and of non-strange D mesons and charged pions
    in $2<\ptrans<3~\gev/c$ (right panel). The filled boxes represent the D-meson systematic uncertainties that are correlated in the different
    centrality intervals.}
  \label{fig:RAAvsNpart}       
\end{figure*}
\section{Analysis and results} 
\label{sec:pp}
Open charm production is measured in the ALICE experiment via an invariant mass analysis of the 
decay channels $\DtoKpi$ (BR=3.88\%, $c\tau=123~\mum$), $\DtoKpipi$ (BR=9.13\%, $c\tau=313~\mum$), $\DstartoDpi$ (BR=68\%, strong decay)
and $\DsToPhipi\to K^{+}K^{-}\pi^{+}$ (BR=2.28\%, $c\tau=149~\mum$) in pp collisions at $\sqrts=2.76$~and~$7~\tev$, p-Pb collisions
at $\sqrtsNN=5.02~\tev$ and Pb-Pb collisions at $\sqrtsNN=2.76~\tev$. A similar analysis strategy was used in the different
collision systems~\cite{DmesonPP,DmesonPP276,AliceDspp,DmesonRAA}, based on exploiting the ALICE high track
spatial resolution in the vicinity of the primary vertex of the collision ($\sim 60~\mum$ at $\ptrans=1~\gev/c$), granted by the Inner Track System
silicon detector (ITS), to identify secondary decay vertices displaced by few hundred $\mum$ from the interaction vertex. In order to further enhance the ratio between the D-meson signal
and the large combinatorial background the measurements of the particle time-of-flight from the collision point to the Time Of Flight (TOF) detector 
and of the specific energy loss in the Time Projection Chamber (TPC) gas are used to identify kaons and pions. The pseudorapidity 
acceptance for tracks reconstructed in the ITS, TPC and TOF detectors is $|\eta|<0.9$. 
 
In the left panel of Fig.~\ref{fig:RpPb}, the black points represent the ALICE preliminary average of prompt non-strange D-meson 
$\RpPb$. As for the measurements performed in pp and Pb--Pb~\cite{DmesonPP,DmesonRAA}, the contribution of D mesons coming from 
B-hadron decays is estimated and subtracted using the cross section of D mesons from B-meson decays calculated
with FONLL~\cite{fonll}, the reconstruction efficiency for prompt and secondary D mesons and using a range of hypotheses for the nuclear 
modification factor of feed-down D mesons ($\RpPbfd$). In p--Pb collisions
$\RpPbfd/\RpPbD=1$ is assumed and the range $0.9<\RAAfd/\RAAD<1.3$ is spanned for assigning a systematic uncertainty. The range was
chosen considering the predictions from calculations including initial state effects based on EPS09 nuclear PDF
parametrizations~\cite{epsZeroNine} and from the model based on the Color Glass Condensate (CGC) framework 
described in~\cite{CGCfujii}. $\RpPb$ is compatible with unity
in the range $1<\ptrans<24~\gev/c$. In the centre of mass system of the p--Pb collisions, the measurement covers 
the rapidity range $-0.96<y<0.04$. Within the current statistical and systematic uncertainties, no evidence of a dependence 
of the production cross section on rapidity is observed within this window, as shown in the right panel of the same figure
for $\Dzero$ in $2<\ptrans<5~\gev/c$, $5<\ptrans<8~\gev/c$ and $8<\ptrans<16~\gev/c$. The measured $\ptrans,y$-differential cross-section is 
compatible with predictions obtained by scaling the y-differential cross-section calculated with FONLL~\cite{fonll} 
by the Pb atomic mass number and by a $\RpPb$ estimated, as a function of $\ptrans$, on the basis of the MNR calculation~\cite{mnr} and 
EPS09 nuclear PDF parametrizations~\cite{epsZeroNine}. The latter $\RpPb$ estimate describes well the measured $\RpPb$~\cite{graziaSQM}. The 
comparison with the preliminary $\RAA$ measured in the 7.5\% most central Pb--Pb collisions in $1<\ptrans<36~\gev/c$, shown
in the left panel of Fig.~\ref{fig:RpPb}, highlights that the suppression observed in Pb--Pb collisions (about a factor 5 for $\ptrans\sim 10~\gev/c$)
is predominantly induced by final state effects due to charm quark energy loss in the medium. 

The preliminary comparison of the centrality
dependence of the nuclear modification factors of non-strange D mesons and of $\jpsi$ from B-meson decay measured with 
CMS~\cite{CMSjpsiPAS}, displayed in the left panel of Fig.~\ref{fig:RAAvsNpart}, represents an indication for a stronger suppression 
of charm than beauty in central Pb--Pb collisions. The $8<\ptrans<16~\gev/c$ range was chosen for D mesons 
\begin{figure*}
  \centering
  \includegraphics[width=6.5cm,clip]{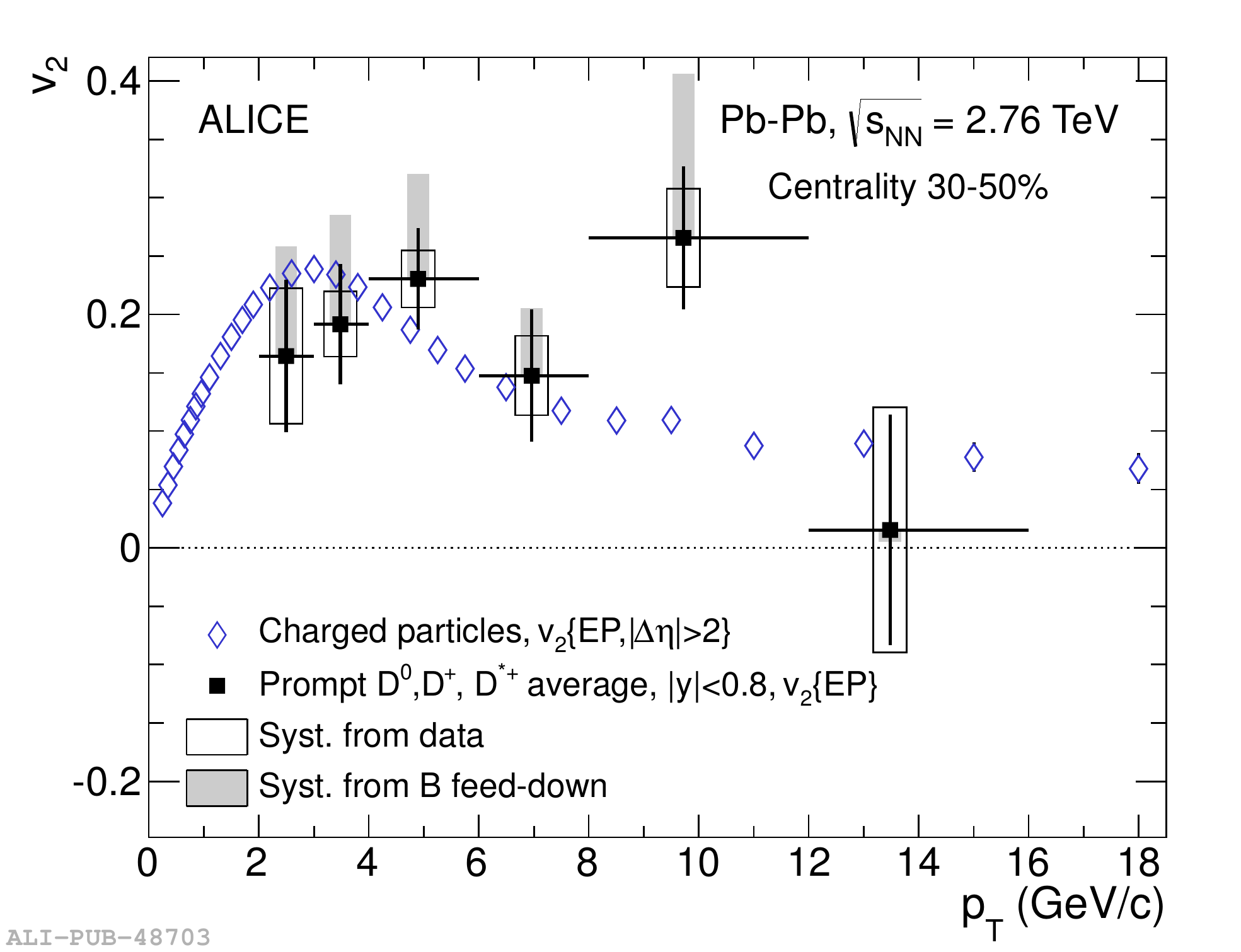}
  \includegraphics[width=6.1cm,clip]{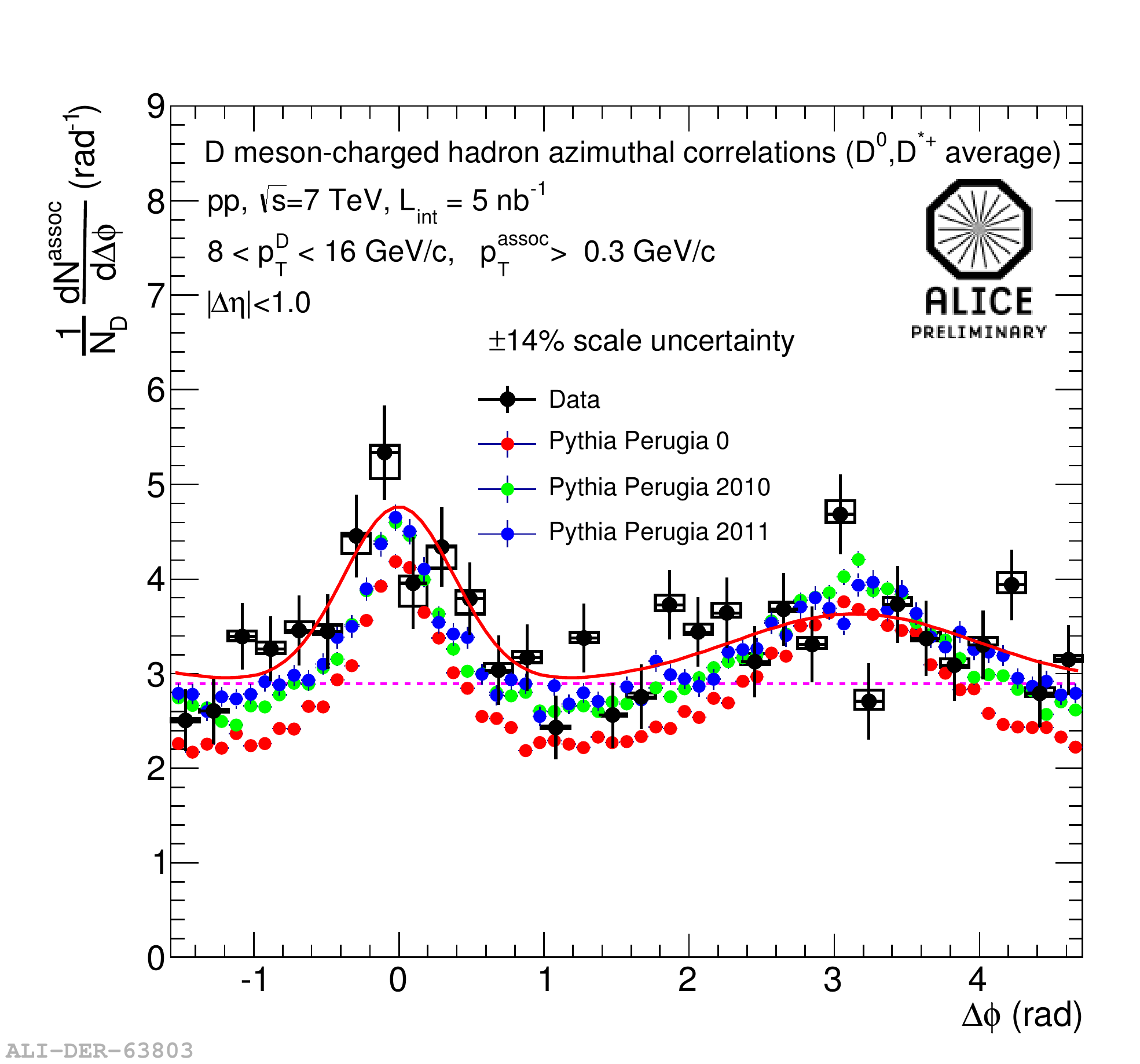}
  \caption{Left panel (from~\cite{Dmesonv2}): comparison between $\Dzero, \Dstar$ and $\Dplus$ average $\vtwo$ and charged 
    hadron $\vtwo$ measured in 30-50\% central Pb--Pb collisions. Right panel: azimuthal correlations 
    between D mesons (average of $\Dzero$ and $\Dstar$ results) and charged hadrons in pp collisions at $\sqrts=7~\tev$, 
    compared to expectations from simulations with the PYTHIA event generator~\cite{pythia,PerugiaTunes}. The boxes represent the systematic uncertainty not correlated
    in $\Delta\phi$.
  }
  \label{fig:v2correl}       
\end{figure*}
in order to have a similar kinematic range than that of B mesons decaying in a $\jpsi$ in the measured $6.5<\ptrans<30~\gev/c$
range. For the subtraction of non-prompt D~mesons from B-meson decays, on the basis of CMS results, $\RAAfd/\RAAD=2$ was assumed as a central value and 
the range $1<\RAAfd/\RAAD<3$ was considered for assigning 
the systematic uncertainty related to this hypothesis. 
As reported in~\cite{DmesonRAA,GrelliQM},
the D-meson $\RAA$ is compatible within uncertainties with the charged-hadron $\RAA$ for $\ptrans>1~\gev/c$ in central
Pb--Pb collisions. The same observation holds for more peripheral collisions, as also visible in the right
panel of Fig.~\ref{fig:RAAvsNpart}, where the centrality dependence of D-meson and charged-pion nuclear modification factors
is shown for $2<\ptrans<3~\gev/c$. 

In the left panel of Fig.~\ref{fig:v2correl} the first measurement of D-meson elliptic flow 
in heavy-ion collisions is shown. The measurement, performed with ALICE in the 30-50\% centrality range~\cite{Dmesonv2}, exploits the event plane method, in which 
the correlation of the particle azimuthal angle ($\varphi$) to the reaction plane $\Psi_{\rm RP}$ is analyzed. The 
reaction plane is estimated via the event plane $\Psi_{2}$, which is obtained
from the azimuthal distribution of a (sub-)sample of tracks in the event~\cite{VoloshinV2}. The measurement
represents a 5 $\sigma$ observation of $\vtwo>0$ in the range $2<\ptrans<6~\gev/c$, with
an average of the measured values in this interval around 0.2. A positive $\vtwo$ is also observed for $\ptrans>6~\gev/c$, which most likely originates
from the path-length dependence of the partonic energy loss, although the large uncertainties do not allow
a firm conclusion. The measured D-meson $\vtwo$ is comparable in magnitude to that of charged particles,
which is dominated by light-flavour hadrons~\cite{hadronV2}. 
This suggests that low momentum charm quarks take
part in the collective motion of the system. 

As an outlook, in the right panel of 
Fig.~\ref{fig:v2correl}, the first measurement of the azimuthal correlations 
between D mesons and charged particles in pp collisions at $\sqrts=7~\tev$ is shown (more details on the analysis technique can
be found in~\cite{ColamariaSQM}). The measurement, which is the 
average of $\Dzero$ and $\Dstar$ correlations with charged particles, is compatible with the expectations obtained 
from the PYTHIA event generator~\cite{pythia} (tunes Perugia~0, Perugia~2010, and Perugia~2011~\cite{PerugiaTunes}), considering the large statistical and systematic uncertainties.
The latter is dominated by the~14\% uncertainty on the normalization. The larger statistics
expected from upcoming runs at the LHC in 2014-2018, should allow for a precise measurement already in run 2 and the upgrade of the ALICE detector 
during the long shut down in 2018~\cite{ALICEupgradeLOI} should give access to this observable also in Pb--Pb collisions with run 3 data.





\end{document}